\documentclass[preprint,journal]{vgtc}       

\ifpdf
  \pdfoutput=1\relax                   
  \usepackage{graphicx}                
  \DeclareGraphicsExtensions{.pdf,.png,.jpg,.jpeg} 
\else
  \usepackage{graphicx}                
  \DeclareGraphicsExtensions{.eps}     
\fi%

\graphicspath{{figures/}{pictures/}{images/}{./}} 

\usepackage{microtype}                 
\PassOptionsToPackage{warn}{textcomp}  
\usepackage{textcomp}                  
\usepackage{mathptmx}                  
\usepackage{times}                     
\usepackage{cite}                      
\usepackage{tabu}                      
\usepackage{booktabs}                  
\usepackage{color}                     
\usepackage[table,xcdraw,dvipsnames]{xcolor}        

\ieeedoi{10.1109/TVCG.2019.2934658}

\onlineid{1034}

\vgtccategory{Research}
\vgtcpapertype{Application/Design Study}

\title{LightGuider: Guiding Interactive Lighting Design using Suggestions, Provenance, and Quality Visualization}

\author{Andreas Walch, Michael Schw\"{a}rzler, Christian Luksch, Elmar Eisemann, and Theresia Gschwandtner}
\authorfooter{
\item
 Andreas Walch is with VRVis Forschungs GmbH. E-mail: walch@vrvis.at.
\item
 Michael Schw\"{a}rzler is with TU Delft. E-mail: m.schwarzler@tudelft.nl.
 \item
 Christian Luksch is with VRVis Forschungs GmbH. E-mail: luksch@vrvis.at.
\item
Elmar Eisemann is with TU Delft. E-mail: e.eisemann@tudelft.nl.
\item
 Theresia Gschwandtner is with TU Wien, Visual Analytics research division. E-mail: theresia.gschwandtner@tuwien.ac.at.
}

\shortauthortitle{Walch \MakeLowercase{\textit{et al.}}: LightGuider}

\abstract{LightGuider is a novel guidance-based approach to interactive lighting design, which typically consists of interleaved 3D modeling 
operations and light transport simulations. Rather than having designers use a trial-and-error approach to match their illumination constraints and aesthetic goals, LightGuider supports the process by simulating potential next modeling steps that can deliver the most significant improvements. LightGuider takes predefined quality criteria and the current focus of the designer into account to visualize suggestions for lighting-design improvements via a specialized provenance tree. 
This provenance tree integrates snapshot visualizations of how well a design meets the given quality criteria weighted by the designer's preferences. This integration facilitates the analysis of quality improvements over the course of a modeling workflow as well as the comparison of alternative design solutions.
We evaluate our approach with three lighting designers to illustrate its usefulness. 

} 

\keywords{guidance, 3D modeling, lighting design, provenance, global illumination}

\CCScatlist{ 
 \CCScat{K.6.1}{Management of Computing and Information Systems}%
{Project and People Management}{Life Cycle};
 \CCScat{K.7.m}{The Computing Profession}{Miscellaneous}{Ethics}
}

\teaser{
  \centering
  \includegraphics[width=\linewidth]{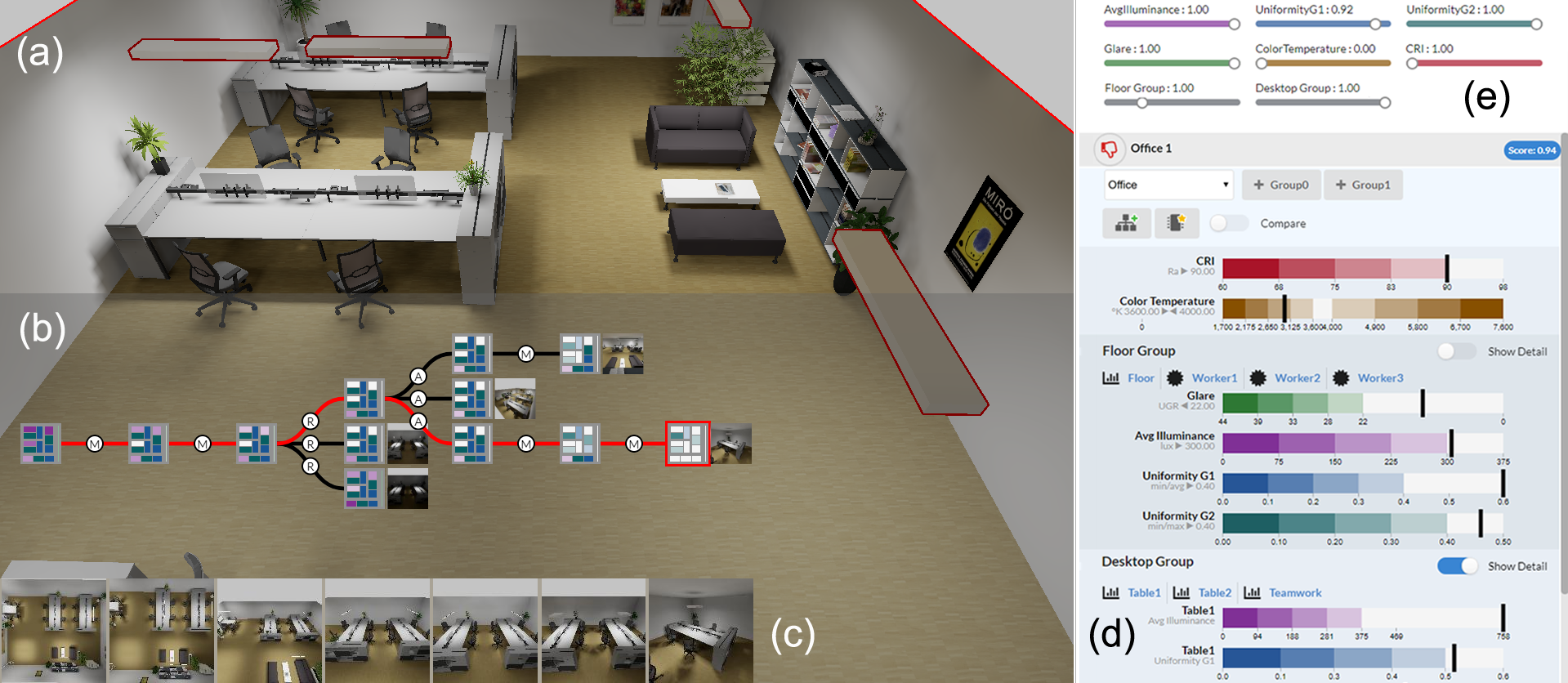}
  \caption{The components of \emph{LightGuider}: (a) a 3D modeling view to place and modify luminaires, augmented with (b) a provenance tree, depicting several sequential modeling steps and parallel modeling branches, integrating information on the quality of the individual solutions, and providing guidance by pre-simulating and suggesting possible next steps to improve the design. A film-strip-like visualization (c) of screenshots helps to depict the evolution up to the currently selected state. A quality view (d) informs about the fulfillment level of the illumination constraints that need to be met, using bullet charts. Changing the weights of these constraints (e), and therefore, the lighting designer's focus, triggers an update of the provenance tree node visualizations (reflecting the weights of the constraints in the distribution of the treemap space). Moreover, the defined weights are also considered in the generation of new suggestions, which are tailored towards satisfying constraints with higher weights.
  }

\label{fig:teaser}
}

\vgtcinsertpkg

\begin{document}


\firstsection{Introduction}

\maketitle

Lighting design is the process of placing luminaires in a 3D environment in such a way that the emitting light fulfills both technical and aesthetic requirements. 
Industry norms or customer wishes define the amount of light with a specific distribution (e.g., uniformity is often a key requirement) to illuminate a certain area (e.g., desktops). Simultaneously, it is important for lighting designers to pay attention to architectural considerations.

In contrast to standard CAD modeling, in which each manipulation typically leads to immediate visual feedback, the process for lighting design is usually \emph{decoupled and indirect}. Simulating lighting in a new scene configuration is costly, because light propagates through a scene by reflecting off many surfaces. Light characteristics, geometry, and surface material properties (that influence the scattering of the light) all affect the outcome, which makes it impossible to predict an exact appearance without an accurate simulation. Consequently, the result of a modeling step is tough to predict, even for experienced designers.


When relying on interactive workflows that approximate the result by relying on recent advances in terms of hardware and algorithms, lighting design takes a \emph{trial-and-error}-based approach. A designer tries to converge to a solution in which all constraints are fulfilled as good as possible. The process is tiresome and the outcome is usually a single local optimum, while all potential alternatives remain undiscovered.
Today's workflows do not support multiple solutions simultaneously, which further reinforces this problem. Furthermore, it is hard to quantify and compare the potential of different solutions during the modeling process.

Approaches to improve the effeciency of lighting design through automation have so far not found their way into today's workflows. One reason is that the parameter space for the possible designs is extremely large. To solve this \emph{multi-objective optimization problem}, an algorithm has to consider all (theoretically infinitely many) spatial configurations of light sources in terms of position and orientation, as well as the possibility to switch or dim them, increase or decrease their number, or to replace them by a different model. These decisions of a lighting designer are input parameters of a complex lighting simulation that needs to fulfill given constraints (e.g., all desktops need to be sufficiently illuminated without glaring persons). Furthermore, it is difficult to integrate aesthetic considerations, in non-interactive, simulation-based approaches. This places our work in the field of visual parameter space analysis according to Sedlmair et al.~\cite{6876043}. Moreover, the process of lighting design involves a stepwise improvement of the design while constantly evaluating if and how an individual decision improved the design and what constraints are affected in which way, as well as the creation and comparison of multiple alternative designs with subsequent fine-tuning of design decisions. Efficiently supporting this process requires an interactive visualization of provenance information including quality criteria for alternative branches of designs.


To support interactive lighting design, we propose to structure the workflow using a Visual Analytics (VA) approach. It combines a 3D simulation of the lighting design with an enhanced \emph{provenance tree}. The tree not only allows lighting designers to keep track of and assess modeling actions, illumination states, and alternative designs, but it also acts as a guidance tool for suggestion-based modeling. The graphical user interface elements of our prototypical system can be seen in \autoref{fig:teaser}. Specifically, our contributions are: 

\begin{itemize}
    \item The first interactive modeling system with automatically-generated design suggestions, while at the same time preserving artistic freedom to enable lighting designers to account for architectural and aesthetic considerations.
    \item A guidance mechanism, which takes the current illumination state and the lighting designer's preferences into account to suggest promising design improvements by simulating multiple alternative next steps. 
    \item A visually-enhanced provenance tree to immediately assess the quality of different lighting designs, monitoring the progress of the design process, and comparing alternative solutions.

\end{itemize}


\section{Related Work}
We structure works related to our novel VA lighting-design solution with respect to different scientific fields.

\subsection{Lighting Design}

Lighting design using commercial tools~\cite{dialux,relux,agi32} is mainly a 3D modeling task (a designer selects, places, and orients light sources), followed by a global illumination simulation. Accurate simulations, with multiple light bounces take at least several seconds (using GPU-based approaches), but sometimes minutes or even hours to complete.
This shows the difficulty of trial-and-error-based methods, whose success for finding a local optimum that fulfills the constraints (such as industry norms) depends strongly on skill and experience of the designer. None of the commercially available tools provide means for comparing solutions, modeling suggestions, or anything more than a linear undo/redo queue. Their big advantage is the complete artistic freedom in terms of parameter selection---an aspect that seems to have a big impact on whether designers choose a certain system. 

In the scientific domain, several approaches automate or simplify light-source placement and orientation---either with procedural methods as suggested by Schwarz and Wonka~\cite{Schwarz2014}, or by ``painting'' the parts of a scene for illumination~\cite{Schoeneman1993,Okabe2007,Shesh2007,Pellacini2007,Lin2013,SLE17}. While these methods deliver solutions to certain aspects, they ignore the iterative, interactive workflow of lighting designers, in which a large variety of considerations (that may not all be quantifiable) play an important role. 
Other approaches focus on interactivity, and try to decrease the feedback cycles between modeling and simulation. Both Luksch et al.~\cite{PB-VRVis-2013-002} and Kr\"osl et al.~\cite{PB-VRVis-2017-018} rely on fast, GPU-based simulations. 
Despite being efficient, they do not offer guided modeling proposals or methods to explore and compare parallel modeling tracks.

Sorger et al.~\cite{sorger-2015-litevis} tackle the problem of comparing different light configurations by linking the simulation results and a spatial view with non-spatial ranking and comparison visualizations. Their idea of setting the importance of certain criteria to compute the overall score (i.e., giving more weight to certain illumination requirements, to certain scene objects, or to global factors like maintenance costs) during the decision process, has influenced our work. Nevertheless, their approach does not take the modeling process itself into account and presumes the availability of a high number of valid, pre-simulated lighting configurations. This assumption rarely holds in real-world scenarios (due to the trial-and-error-based methodology converging to a single valid solution), raising the need for novel methods that produce multiple solutions in parallel. Other solutions, such as proposed by Simons et al.~\cite{SHPRALDEE18}, record light rays and offer visual-analytics tools to explore, evaluate, and compare light interactions, potentially involving several scenes. Nevertheless, they do not offer suggestions for scene manipulations to fulfill given constraints. 

\subsection{Visual Parameter Space Analysis}

The modeling interactions of LightGuider, such as placement of luminaires or rearrangement of the scene, are the input parameters of a complex lighting simulation, in which lighting designers try to generate output values, such as glare or direct illumination, that adhere to certain constraints. Many solutions rely on an exhaustive sampling of the parameter space, such as Bruckner and M\"oller~\cite{brucker-2010-RES} who pre-compute particle simulations and then cluster visually similar results. For our use case it is not possible to conduct a sufficiently dense sampling in reasonable time, therefore, approaches relying on pre-computation are not applicable. Coffey et al.~\cite{coffey2013design} propose a tool for the simulation-driven design of biopsy needles. Mechanical engineers pick a specific design and evolve variations from that by dragging interactions. Although they still rely on pre-computed design results, their workflow is similar to LightGuider's. 
Flood management and barricade design typically deals with a large number of input and environmental parameters. The decision support system Visdom~\cite{6280550,Waser2010} employs a tree visualization that allows domain experts to evaluate the quality of barricading plans, however, quality in this context can be quantified in damages and water levels, while LightGuider needs to convey the fulfillment of complex design constraints and aesthetic properties to light designers. Mark et al.'s \emph{Design Galleries} approach~\cite{Marks97} uses thumbnails of images or animations computed in a preprocessing step by sampling a broad range of input parameters. Based thereon, Pfister et al. \cite{Pfister:2001:TFB:616070.618820} apply this technique for the selection of transfer functions in volume rendering. 

In accordance with Sedlmair et al.~\cite{6876043}, we classify LightGuider as follows: Utilizing an interactive lighting simulation, lighting designers start out with a single sample and generate new \emph{samples} on-the-fly supported by guidance mechanisms (see \autoref{sec:relworkguidance}) suggesting alternatives in the parameter space. Immediate feedback of the simulation results provides them with \emph{local-to-global} navigation. As lighting designers need to evaluate qualitative as well as quantitative aspects of the simulation output, the domain goals of LightGuider present a mixture of \emph{design} and \emph{engineering} domain goals, which are reached through the \emph{optimization} of both. As a secondary analysis objective we identify \emph{partitioning} in the elaboration of alternative designs to illustrate different trade-offs.

\subsection{Provenance Visualization}
When it comes to provenance information in visualization, Ragan et al.~\cite{ragan:tvcg:2016} give a comprehensive overview of the different types of provenance information (e.g., the history of data editing, the history of graphical views and visualization types, or the history of interactions) and different purposes of using them in the context of visualization (e.g., recall different states of the analysis, action recovery, or collaboration). 
However, there are varying approaches to visualize this information. The most common choice is presenting the provenance tree as a node-link diagram that shows the sequence of states and alternative branches of a workflow as described by Simmhan et al.~\cite{simmhan_survey_2005}. 

Stitz et al.~\cite{stitz:CGF:2016} focus on the scalability of node-link diagrams for encoding a history of analysis workflows. They use filtering, node aggregation, as well as a user-interest driven expansion of nodes (i.e., a degree-of-interest function) to make the tree more comprehensible. In a different work, Stitz et al.~\cite{2018_vast_knowledge-pearls} use a provenance tree for visualizing automatically recorded user interactions and visualizations. Again, they focus on the efficient retrieval of analysis states by offering different possibilities for querying the data (e.g., query by user-generated examples). These works offer sophisticated solutions to scalability problems of provenance trees in the form of node-link diagrams, as well as solutions for efficient interaction with large trees. However, they do not focus on integrating visual representations of additional information for each tree node. Our application scenario requires a quick visual comparison of multiple numerical variables (i.e., illumination constraints) for each state to enable the assessment of changes of quality for each lighting design action as well as trends of the lighting design process and of alternative workflows.

Bors et al.~\cite{bors:eurovis:2018} visualize provenance information for data quality management. They visualize the history of data editing actions in a node-link diagram augmented with visual information of how many data records have been deleted or added at each step. 
Furthermore, they combine this node-link diagram with a visualization that shows how the data quality changes over time, giving details of the amount of data records that violate different specified data quality metrics. While they pair the provenance tree with an additional visualization (i.e., a stacked bar chart) of multiple numerical variables, this can only be done for one branch of the tree at a time, and thus, the comparison of multiple alternative workflows is not supported. 

Besides node-link diagrams, there are examples of other visualization types used to show provenance information. Vi{\'e}gas et al.~\cite{viegas:chi:2004}, for instance, visualize the history of the editing that was applied to a Wikipedia (wikipedia.org) page in a flow-like visualization.
This visualization is specifically designed to represent one page with text running from top to bottom, but the only interactions it supports (indirectly) are ``adding text'' and ``removing text''. Thus, it does not lend itself to our problem scenario.
Another approach by Su et al.~\cite{sara:mit:2009} shows the editing history of illustrations. They provide a superimposed visualization of two illustration states with ``before'' states rendered semi-transparent. Moreover, the illustration is augmented with arrows, icons, and color. Arrows and icons indicate spatial transformations of (parts of) the illustration, while color indicates user changes. This is a specialized design for the problem at hand and cannot be transferred to our application scenario. 

\subsection{Guidance}
\label{sec:relworkguidance}

Guidance in visualization as defined by Ceneda at al.~\cite{ceneda:tvcg:2017} can be found in various forms and application scenarios. However, only a few approaches relate to our problem at hand.
Bouali et al.~\cite{Bouali:vc:2016} present a guidance approach to automatically generate a set of information-visualization designs appropriate for the given data and tasks. A selection of the most useful visualization mappings is input to the guidance mechanism, and influences future suggestions. 
O'Donovan et al.~\cite{O'Donovan:CHI:2015} present a similar approach that helps in creating graphic design layouts. The system interactively suggests changes in the position, scale, and alignment of elements that are placed on a page. 
Both systems present guidance approaches to optimize a design. 

Yang et al.~\cite{Yang:exn:2007} present a guidance approach that helps to discover interesting data and patterns based on the system user's interests. They provide a system to extract, combine, refine, and visualize such findings of interest. They distinguish between user-driven and data-driven findings. In our work we combine user-driven (i.e., the user chooses which areas and which illumination constraints are more important than others) with data-driven (i.e., optimizing the current design with respect to specified illumination constraints) steering of the guidance suggestions.

\section{Background}
\label{sec:background}
In this section we define important concepts in the context of lighting design.
\subsection{Generating Illumination Data}
\label{sec:generatingillumination}
\begin{figure}[ht]
    \centering
    \includegraphics[width=0.95\linewidth]{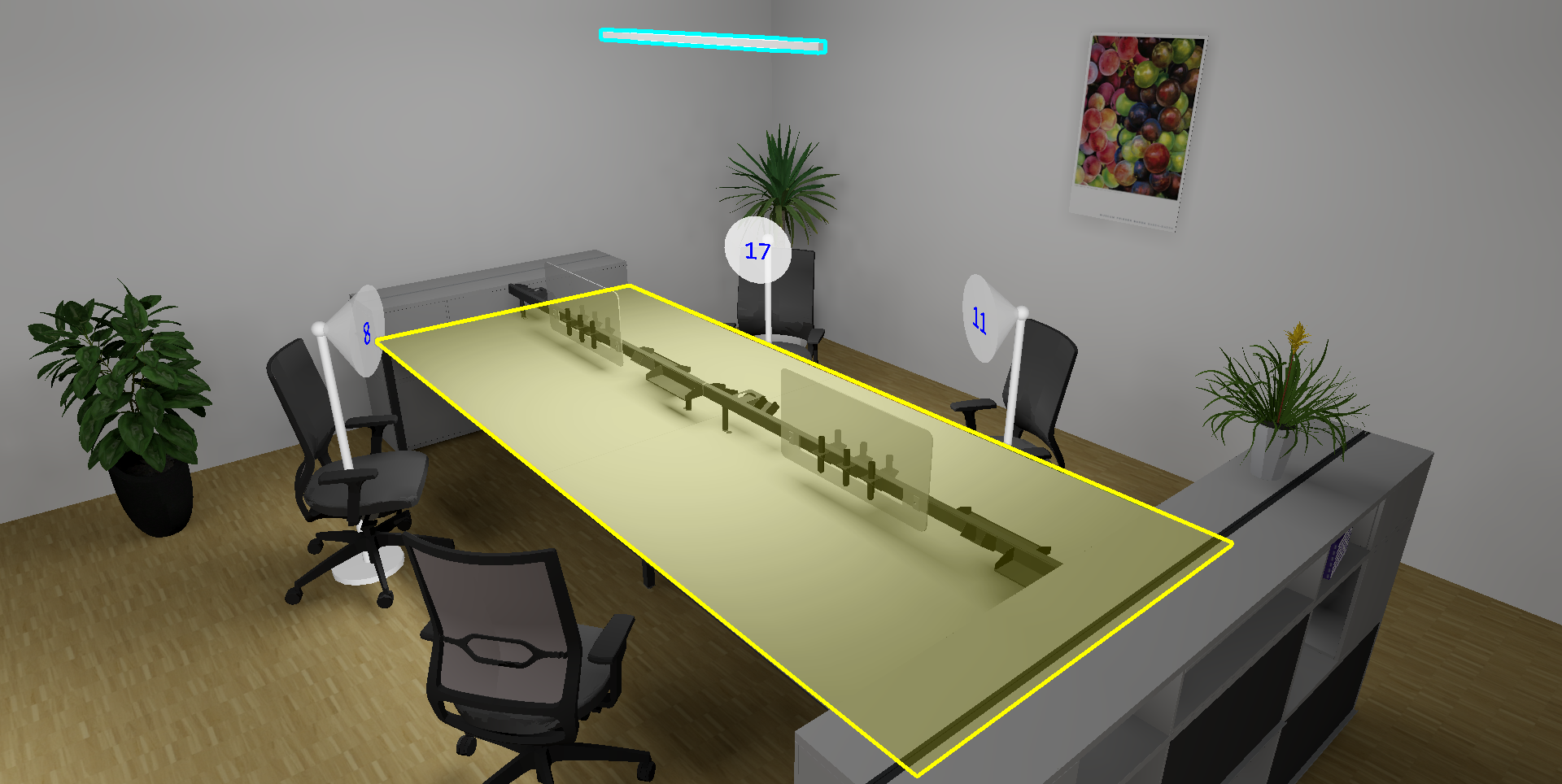}
    \caption{A \emph{measuring surface} on the desktop, and \emph{glare probes} (simulating a person's field of view while working on this desk) are used to gather illumination information, such as the average illuminance and the uniformity on an area, or if a person is likely to be blinded. Depending on the semantics, the target values of the measurement objects are set according to industry norms. Only designs in which all of these target values are reached are considered valid. To fully support linking and brushing in case of occlusions, outline-based highlighting enhancements and semi-transparent area overlays are rendered without depth-testing.}  
    \label{fig:measureSurface}
\end{figure}

Global light transport simulations give an estimation of the propagation of light from the sources 
via several bounces off scene surfaces 
to each point in the scene. 
A realistic visualization of the light distribution in a 3D space helps to get an aesthetic impression of the result, while the quality of a lighting design is typically illustrated using a false color coding. 
Lighting designers face many regulations and constraints given by industry norms and customers, such as specific regions that have to be illuminated with a minimum amount of light and a uniform distribution, or certain critical positions for which no blinding should occur (e.g., a person sitting at a desk). 
For this reason, designers rely on \emph{measuring surfaces} (measuring incoming light in this area) and \emph{glare probes} (simulating a person's field of view from a certain point and direction) that are placed in the scene (see \autoref{fig:measureSurface}). 
They are typically linked to scene objects that have a certain semantic meaning (e.g., a desk
) with specific norm-based target values assigned to them (see \autoref{illuminationGoals}). 
In our approach, the light transport simulation is performed using a GPU-based many-light approach similar to Luksch et al.\cite{PB-VRVis-2013-002},
and the incoming light is stored in the texels of a \emph{light map} that covers all surface geometry.

\paragraph{Input Parameters.}
Similar to Sorger et al.~\cite{sorger-2015-litevis} and according to the classification by Sedlmair et al.\cite{6876043}, we distinguish between \emph{control parameters} (the lights in the scene, including their position and orientation), \emph{model parameters} (defining the accuracy of the simulation, e.g., the resolution of the light maps, or the number of bounces) and \emph{environmental parameters} (the scene with its materials, additional illumination by sunshine / weather, etc.) as inputs for our simulation. In our work we focus solely on the modification of the control parameters. 
As a simulation run is always coupled to a certain configuration of light sources modeled by the designer or proposed by our guidance system, we refer to them together as \emph{modeling state} or \emph{illumination state} in this paper.

\subsection{Illumination Constraints}
\label{illuminationGoals}

Apart from aesthetic considerations, we use the following quantifiable, industry-established metrics (e.g., as described in Zumtobel's Lighting Handbook~\cite{lightinghandbook}) to rate the quality of a solution:

\paragraph{Local indicators.}
The semantics of objects or regions in the scene, e.g., a desktop, are linked with local indicators. By assigning measuring surfaces, statistical aggregations of the individual measurements (i.e., of all the texels of the measuring surface) can be generated to define the illumination quality. We focus on the industrially most-widely used indicators, namely \emph{Average Illuminance (AVG)}, \emph{Uniformity G1} (= min/avg), and \emph{Uniformity G2} (= min/max).

\paragraph{Person-oriented indicators.}
Contrary to Sorger et al.~\cite{sorger-2015-litevis}, we also take a view-dependent criterion into account---the \emph{Unified Glare Rating (UGR)} defined by the CIE~\cite{CIE_117-1995-1}, giving an estimation of whether a person will be blinded by the lights in the scene. The optimal placement and configuration of the glare probes (simulating a person's field of view) is not as well-defined as for measuring surfaces, because of the theoretical possibility of infinitely many probes. The designer chooses a feasible number at relevant spots. 





\paragraph{Global indicators.}
We use the meta data of the luminaires to derive global indicators, which are not measured in the simulation, nor linked to certain objects. We consider \emph{Color Temperature} in Kelvin (K) by computing the weighted average (per Lumen) of all light sources in a group, and the \emph{Colour Rendering Index} (CRI), for which we take the minimum value of the corresponding lights. 

According to the DIN Standard EN 12464-1~\cite{DIN_EN_12464-1} or customer requirements, the target values for these indicators are set per measuring surface, glare probe, or the whole scene, and are referred to as \emph{illumination constraints}. We consider a solution valid if all of these illumination constraints are reached. 
Note that subjective opinions (such as aesthetics) could still disqualify a valid solution.

\subsection{Interactions and Tasks in Lighting Design}
\label{sec:tasks}

The actions that a lighting designer traditionally performs to find a suitable solution fulfilling the illumination constraints after setting up the scene and defining the target values can coarsely be described by the following workflow: 

\begin{enumerate}
    \item Selection, placement, and alignment of luminaires, followed by a simulation run
    \item Verification of all illumination constraints followed by the designer's choice of which requirement to improve 
    \item Selection of a modeling action (e.g., change the height of a light) that is likely to improve the selected illumination requirement and maintains the aesthetic goals, followed by a new simulation run
    \item Repetition of the two previous steps until all illumination constraints are fulfilled and the aesthetic expectations are met
\end{enumerate}

This iterative approach typically leads to the evaluation of only ``one path to a valid solution''.
Only in cases in which the customer is not satisfied, further, alternative solutions are generated. Moreover, it should be noted that since the different illumination constraints are connected, a modeling action would usually not only improve the targeted constraint, but might also have negative effects on other constraints (e.g., lowering a light would improve the \emph{Avg. Illuminance} on a desktop but also enhance \emph{Glare}). 

\section{LightGuider Design Requirements}

We designed LightGuider to address the trial-and-error work-flow of lighting design, by automatically generating new solutions and visually guiding the lighting designer to find an optimal solution. Moreover, we preserve the creative procedure of lighting design by enabling the designer to try and to compare different directions and consider aesthetic aspects in addition to numerically defined illumination constraints. 

In the course of more than ten years of tight collaboration with lighting designers, we identified the following aspects required to improve the lighting-design process: 
\begin{itemize}
    \item[R1] Lighting design should be \emph{interactive} to account for artistic preferences. 
    \item[R2] A \emph{3D simulation} of current and previous lighting designs are important to evaluate aesthetic considerations.
    \item[R3] The trial-and-error-based methodology should be superseded by a good \emph{visual overview of the history} of modeling operations, parallel design branches, and intuitive restoration of modeling states. 
    \item[R4] A fast \emph{visual summary of the quality} of each state (i.e., how well it meets the illumination constraints) is required to \emph{evaluate} the impact of certain modeling operations, the tendency of improvement within a modeling workflow, different intermediate solutions, as well as parallel design branches. 
    \item[R5] Visual means for a \emph{fast comparison} of these states and solutions should be provided. 
    \item[R6] The multi-objective optimization problem of finding promising light configurations within a huge parameter space should be \emph{guided by automatic means}.
\end{itemize}

\section{LightGuider Visualization Design \& Interactions}
To make informed decisions about our visualization and interaction design we analyzed (1) the data we need to communicate, (2) the intended users of the approach, as well as (3) the tasks they have to solve (as described by Miksch and Aigner~\cite{Miksch2014}). The data consists of multiple aspects: the scene for which a lighting design should be found together with placed light sources, the outcome of light transport simulations (see \autoref{sec:generatingillumination}), the illumination constraints that should be met (see \autoref{illuminationGoals}), as well as the history of modeling operations. Our VA approach is specifically developed for lighting designers, who are our expected user group. These lighting designers need to choose a set of multiple lighting-design modeling parameters to find a lighting solution that fulfills not only their aesthetic requirements but also a number of illumination constraints. 

\subsection{System Overview} 
\label{sec:systemoverview}

LightGuider is an enhancement to a lighting design prototype focused on interactive editing of a 3D scene (R1) coupled with a fast light simulation (see \autoref{fig:teaser}). It consists of a spatial 3D editing, simulation and visualization view (see \autoref{sec:3dview}), which is partly and transparently overlaid by the core component of our novel system---a graph-based 2D visualization that is used to depict the modeling progress and to guide lighting designers to reach their goals (see \autoref{sec:provenancetree}). 
Integrated into the existing user interface of the software, additional panels give detailed information on a modeling state (see \autoref{sec:bullet}) and allow for adjusting the importance of illumination constraints and measuring surfaces (see \autoref{sec:weighting}). All of these components are connected with regards to linking and brushing. 

We tightly integrated our LightGuider approach into an established and well-evaluated lighting design tool by Luksch et al.~\cite{PB-VRVis-2013-002}, 
which supports placing of light sources, setting their parameters, and evaluating light transport simulations in a 3D scene, and thus, avoiding the need for the designers to learn a new environment. This also allowed us to evaluate whether our novel attempt to guide the workflow was accepted by the domain experts, or whether they would rather prefer their traditional methods while solving the given tasks (see \autoref{sec:evaluation}). 

\begin{figure*}[htb]
    \centering
    \includegraphics[width=\linewidth]{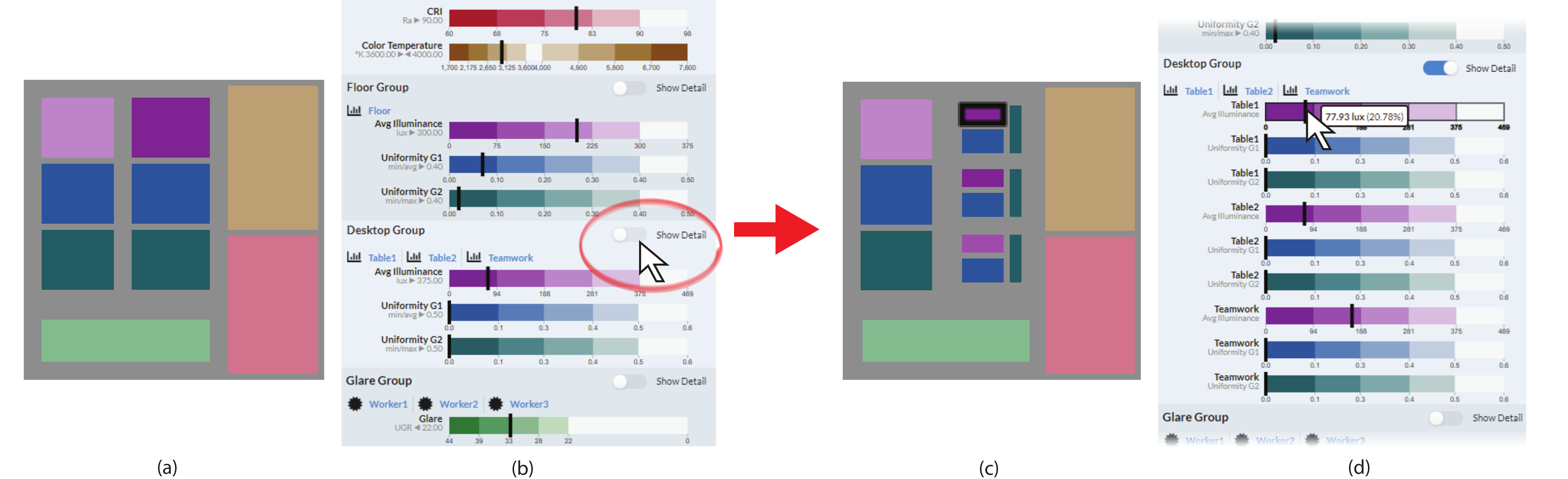}
    \caption{The fulfillment levels of the illumination constraints of a modeling state, represented both as a treemap (see \autoref{sec:qualityvis}) and using bullet charts (see \autoref{sec:bullet}). In the summary-view (a and b), the worst illumination values of a group are depicted. When requesting more details, the information for all measuring surfaces is displayed (c and d). Linking \& brushing and tooltips help to easily grasp shortcomings of the modeling state.}
    \label{fig:tree}
\end{figure*}

\subsection{3D View}
\label{sec:3dview}

The freely navigable 3D view (see \autoref{fig:teaser} (a)) is a combined visualization, editing, and simulation interface. It displays the illuminated 3D scene using the simulation information stored in the light maps in real-time.
Using typical manipulation tools known from CAD tools, light sources (i.e., the simulation's control parameters) can be arbitrarily placed, moved, rotated, removed, exchanged, dimmed, and grouped. 
After each modification, the simulation is immediately restarted asynchronously, continuously converging to a new result while the designer continues to freely interact with the scene. 
Furthermore, measuring surfaces and glare probes can be placed and configured according to the requirements. 
The visualization of measuring surfaces within their real 3D spatial context, i.e., rooms with arbitrary layout containing furniture, such as desks and plants, go along with cumbersome navigation and occlusion impeding visual tasks as classified by Elmqvist and Tsigas~\cite{Elm2008}. We counteract these phenomena using the VISAR methodology suggested by Ortner et al.~\cite{Ortner2016}, as follows: To support lighting designers in the \emph{localization} of, for instance, poorly performing measuring surfaces, a finding command triggers an automated camera animation towards the desired object. 
For \emph{visual discrimination}, we display colored outlines around selected objects, which are rendered without depth-testing, conveying location, shape and orientation even when they are occluded (see \autoref{fig:measureSurface}). 
While this navigatable 3D view was already supported by the used lighting design tool~\cite{PB-VRVis-2013-002},
the remaining views and overlays are specifically designed to support the visual guidance provided by LightGuider.

\subsection{Quality View}
\label{sec:bullet}
For the currently active modeling state displayed in the 3D view, we provide a bullet chart for each illumination constraint that needs to be met (see \autoref{fig:teaser} (d) and ~\autoref{fig:tree} (b)). Each constraint is associated with a unique color. Dark and saturated colors mean that the current design is far from meeting this constraint, while light colors mean that the solution is close. Thus, highly visible colors suggest the need for further fine-tuning of the design, while the absence of color does not attract attention and thus signals that there is no need for further action. 

\paragraph{Color Coding.}
In our application scenario we have six illumination constraints that measure the quality of the lighting design (see \autoref{illuminationGoals}). Mapping the sequential scales of these metrics to color poses different challenges: it requires (1) six well distinguishable color hues, (2) six sequential color scales of these hues (i.e., from white to dark color), and (3) scales with equal brightness values on all levels (i.e., a 30\% quality issue of \emph{Glare} should be perceived equally strong as a 30\% quality issue of \emph{Color Temperature}). To this end, we selected six color scales from ColorBrewer~\cite{colorBrewer} and adapted brightness values where necessary. Considering that two of the six illumination constraints, i.e., \emph{Uniformity G1} and \emph{Uniformity G2}, both describe the uniformity of the light on a measuring surface, we chose similar color hues for those two metrics (see \autoref{fig:tree}). 

\subsection{Provenance View}
\label{sec:provenancetree}
To support the designers in keeping track of their actions and lighting states, we provide a provenance tree as an overlay of the 3D view (see \autoref{fig:teaser} (b)), which enables them to try different alternative designs, jump back and forth between design states, and analyze which actions have which impact on the quality of the design (R3). Each node of the provenance tree represents a fully simulated illumination state.
The different actions of a modeling workflow are indicated by letters: manual (M), add light-source (A), remove light-source (R), dim (d), height move (H), and change light-source (C). While manual (M) stands for any change done by the lighting designer, the other actions indicate changes that have been created by our automatic suggestion system. When hovering the link, a description of the applied action together with its parameters are shown as details on demand.

Selecting a node shows the path to this design state highlighted in red. When adding nodes to the graph, we use smooth transitions to ensure object constancy, which allows the lighting designer to understand how the graph changes. Re-arranging the graph without animations would lead to confusion. Moreover, we enhanced the provenance tree with summary visualizations for each state of the lighting design, as described in the following section.

We opted for a slightly opaque gray overlay for the provenance view to avoid the occlusion of large parts of the 3D scene, while at the same time guaranteeing the continuous readability of the provenance tree while navigating through the 3D scene with changing brightness levels. 

\subsubsection{Quality Visualization of Modeling States}
\label{sec:qualityvis}

\begin{figure*}[ht]
    \centering
    \includegraphics[width=\linewidth]{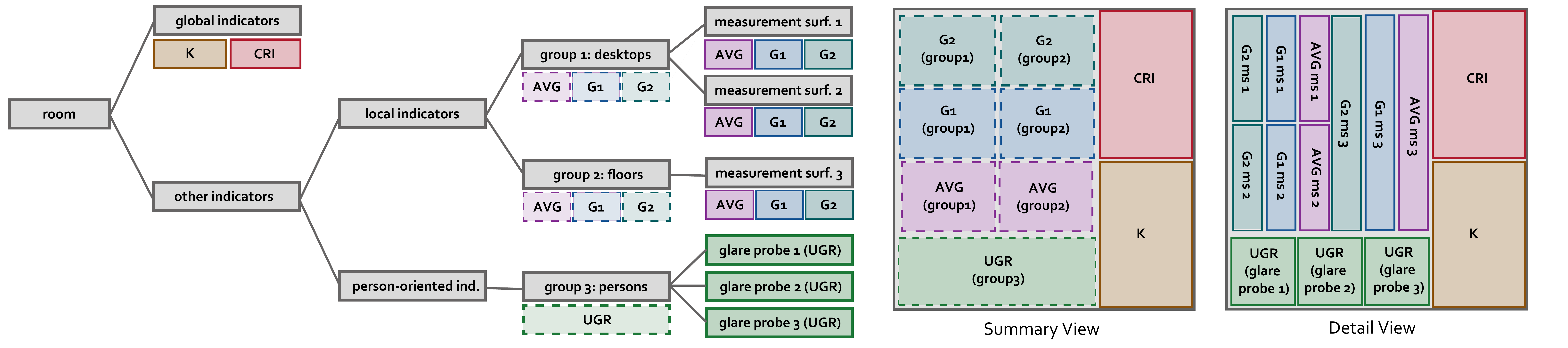}
    \caption{A simple, exemplary hierarchy of illumination constraints in a scene, mapped to the treemap layout. Measurement objects can be grouped into semantic groups by the designer, such as groups with measuring surfaces for desktops and floors, and a general group for glare probes. In the summary-view of the treemap, the worst values of the groups are visualized, and each color occupies a sixth of the treemap space (distributed over the groups). In detail-view, which can be activated for each group separately, the values for the individual measurement objects are displayed.}
    \label{fig:treemap}
\end{figure*}

To visualize the fulfillment of the illumination constraints for each step in the design process, a \emph{treemap} visualization (see \autoref{fig:tree} (a)) is integrated into each node of the provenance tree. Each constraint is associated with a distinct color. 
The color value and saturation describe how good a state fulfills the targeted constraints (R4). This links the treemap visualization to the bullet charts in the quality view (see \autoref{fig:tree} (b)), which also serve as a color legend.

The illumination constraints are hierarchically clustered (see \autoref{fig:treemap}), because some apply for the whole room and others need to be met for each important measuring surface that was defined beforehand, such as a desktop. Thus, some kind of tree visualization is a natural choice to preserve this hierarchical structure.
We chose a treemap visualization because alternative visualizations of hierarchical data with an aspect ratio suited to integrate them into the nodes of the provenance tree, such as a sunburst tree visualization as explained by Stasko et al.~\cite{stasko:2000}, would have dedicated too much space to non-leaf elements such as the grouping of defined measuring surfaces or the whole room. 
This would result in less space for leaf nodes, which is especially detrimental as the treemaps need to be quite small to be integrated into the provenance tree and it is these leaf nodes that encode the important information by color (i.e., how well a specific constraint is met). 
While it can be difficult to read the exact hierarchical nesting from a treemap, we disregard this aspect because information such as hierarchical depths are not important in our scenario. The spatial grouping of semantically similar objects (e.g., all desktops), however, is relevant because the area taken by a group represents its importance (see \autoref{sec:weighting}). 

Furthermore, we provide a summary-view and a detail-view treemap layout (see \autoref{fig:treemap}). 
The summary-view layout shows just the worst value of all measurement object values for each group, i.e., the highest UGR value as well as the lowest values for Avg. Illuminance, Uniformity G1, and Uniformity G2 (see \autoref{fig:treemap}, middle).
A group, definable by the designer, usually contains multiple measuring surfaces or glare probes. 
All six illumination constraint types initially take up the same amount of space in the treemap, i.e., one sixth of the treemap space (in accordance with feedback from domain experts). Given that some illumination constraints can appear in multiple measuring surface groups (e.g., Avg. Illuminance), this sixth is further distributed over these groups (e.g., one twelfth for the desktop group and one twelfth for the floor group).

The detail-view layout shows all glare probe values and all values for each measuring surface within a group (in contrast to showing only the worst value in the summary-view layout; see \autoref{fig:treemap}, right). The detail-view can be activated for each group separately.
In subsequent analysis, the space distribution of the treemap visualizations resemble the weights that are interactively assigned to illumination constraints by the lighting designers (see \autoref{sec:weighting}).

\subsubsection{Comparing Modeling States}
\label{sec:comparisons}

\begin{figure}[htb]
    \centering
    \includegraphics[width=0.7\linewidth]{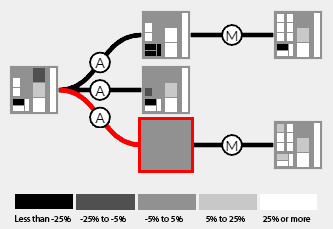}
    \caption{The global comparison visualization of the provenance tree (see \autoref{sec:comparisons}) uses a grayscale color scheme to get an overview on the quality differences of different design states with respect to the currently selected node (in red). Its fulfillment levels of illumination constraints act as the reference, and are visualized in medium gray. Illumination constraints that are fulfilled better in other solutions are visualized lighter, and the ones that are fulfilled worse are displayed darker. }
    \label{fig:comparing}
\end{figure}

To visually support the comparison of different modeling states (R5), the designer can switch the provenance tree visualization to a global comparison mode. This alters the color scheme of the treemaps that are integrated into the nodes of the provenance tree to grayscale (see \autoref{fig:comparing}). The selected node acts as the reference, and all other nodes encode the difference to it. If illumination constraints are fulfilled better/worse, the respective treemap cell is lighter/darker than in the reference node. We decided against a diverging color scheme, because the original color scale of the treemaps trains the lighting designers to interpret lighter/darker colors as better/worse fulfillment of the illumination constraints. By using a grayscale, we preserve this mental model but at the same time avoid confusion with the general color scheme of illumination constraints.
This comparison mode can also be activated for individual comparison of two modeling states by selecting one node and hovering another node (local comparison mode), while the remaining provenance tree preserves the general color scheme.

\subsection{Screenshot Thumbnails View}
\label{sec:thumbnails}

Based on the currently selected modeling state in the provenance tree, and the highlighted ``path'' that was taken through the tree to get there, a film-strip-like arrangement of screenshot thumbnails of the individual modeling steps leading to this state is visualized below the tree (see \autoref{fig:teaser} (c)). The screenshots are taken from the designer's perspective after a change in the scene, when all direct light and some indirect light has been distributed by the simulation (R2).

With this heuristic, designers are able to immediately recall what their actions were, where they took place, and how the resulting illumination looked, making specific previous modeling operations easier to find. The usefulness for comparing modeling states, especially in parallel design branches, is limited though, because the changes in the viewing angle during the interactive modeling procedure lead to large variations. We therefore decided to only display the thumbnails for the linear path through the tree to the current modeling state, and propose other means for comparisons (see \autoref{sec:comparisons}). 

Additionally, we also add thumbnails as quick orientation help at all leaf nodes (i.e., at the current end of the parallel modeling branches). This is especially helpful for presenting modeling suggestions (see \autoref{sec:guidance}), for which we not only show their tree-map nodes, which communicate their quality according to the fulfillment of quantifiable illumination constraints, but also the corresponding screenshots taken from the same viewing angle as the screenshot of the previous step (see \autoref{fig:weights}, right). This way, the designer also gets an impression on the aesthetic impacts of a suggestion and can accept it based thereon.

\subsection{Focus Setting View}
\label{sec:weighting}

\begin{figure*}[htb]
    \centering
    \includegraphics[width=0.95\linewidth]{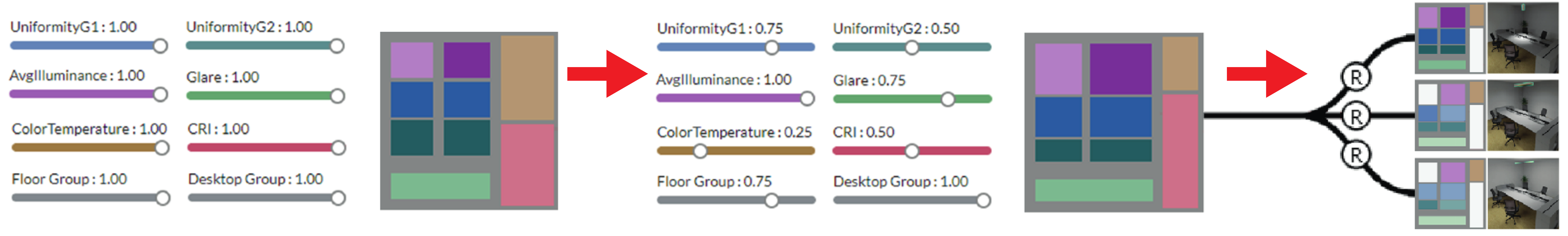}
    \caption{Changing the weights of illumination constraints or group importance changes the layout of the treemaps, putting currently relevant aspects in focus (left; see \autoref{sec:weighting}). The weight change has a global impact, i.e., all nodes in the provenance view share the same treemap layout for an intuitive comparison. Whenever a weight change is performed, new suggestions are computed and displayed based on the weighting (right; see \autoref{sec:guidance}). Here, three replacement actions (R) lead to better results, and are enhanced with screenshots of the new scenarios (see \autoref{sec:thumbnails}).}
    \label{fig:weights}
\end{figure*}

In the \emph{focus setting view}, we provide sliders (see Region (e) in \autoref{fig:teaser} and \autoref{fig:weights}) to set different weights, not only for illumination constraints, but also for user-defined groups of measuring surfaces (e.g., all desktops). Thus, the lighting designer can individually define the importance of each constraint, or the importance of different measuring surface groups (e.g., desktops are more important than the floor). To support linking the sliders with the illumination constraints, we use their assigned colors. 
When giving more weight to a specific constraint or measuring surface group, the corresponding area in the treemap nodes of the \emph{provenance view} is enlarged. Thus, this focus is reflected by the visual representation of the quality of a modeling state (i.e., more important aspects are represented larger and are thus better visible, while less important aspects are smaller). Moreover, the automatically generated lighting design suggestions provided by our guidance system consider these weights when simulating lighting design improvements, as described in the following section.

\section{Generating Lighting Design Suggestions}
\label{sec:guidance}

The purpose of our guidance system is to present the designer suggestions that point to promising new modeling options. We employ our simulation as a useful tool to evaluate potential design decisions in advance, and present them to the lighting designer (R6). This is done whenever a new modeling state is added to the provenance tree, or when the weighting is changed. The suggestions are added as new tree nodes after the current modeling state, and corresponding screenshots give a visual preview (see \autoref{fig:weights}, right). 

To provide meaningful suggestions within a feasible time, two aspects have to be considered: First, the parameter space has to be limited (to reduce simulation time). Secondly, the suggestions have to consider the focus of light designers (to support them in their current task).
For these reasons, we have identified several typical modeling actions a lighting designer performs to improve the fulfillment of certain illumination constraints. Each of these actions can be represented as a small change in the parameter domain (i.e., usually only one or two parameters have to be changed to perform a certain modeling action). By limiting the suggestion choice to the actions that most likely improve the currently focused illumination constraints, the parameter space can be limited, and the suggestions fit the current design goals. In \autoref{sec:provenancetree} we briefly mentioned the basic actions as part of the link labels. In particular, we provide the following actions:

\textbf{Add light:} inserts a light into the scene.
Additional lights can be helpful if the local indicators AVG, or the uniformity indicators G1 and G2 are not satisfied. 
\emph{parameters: position, light-type}

\textbf{Remove light:} deletes a light, chosen either by its ID, or randomly. In case of bright lights within the field-of-view of a person-oriented indicator, the removal improves the UGR. \emph{parameter: light-ID (optional)}

\textbf{Dim lights:} reduces the power of all lights, which is a useful indicator for the designer (and for a future decision algorithm) to reduce the number of lights or to switch to a less powerful (cheaper) version of the current light source model. \emph{parameter: dim factor (0\% - 100\%)}

\textbf{Change height:} shifts one or all lights up or down. This action is only valid for pendant lights. Lowering the lights increases the average illumination but can impair the uniformity, and vice versa. \emph{parameters: shift vector, light-ID (optional)}


\textbf{Change light:} replaces lights by another version (e.g., a stronger / weaker version of the light source), or by switching to another light source collection. A collection change has a drastic impact on the lighting design, and is usually only considered when the global indicators require this.
\emph{parameters: collection-ID or light-ID}


In collaboration with lighting designers, we defined the usefulness of each action for reaching a certain illumination constraint (see \autoref{tab:weights}). We tackle this multi-criteria decision making problem by using a \emph{Weighted-Sum Model (WSM)}~\cite{10.2307/168461}.
For each alternative $A_i$ (the modeling actions, \autoref{tab:weights}~rows) and each criterion $C_j$ (the illumination constraints, \autoref{tab:weights}~columns) a performance value $a_{ij}$ is set.

\begin{table}[ht]
\caption{The unweighted performance values $a_{ij}$ describing the usefulness of modeling actions to improve a specific illumination constraint.}
\label{tab:weights}
\centering
\begin{tabular}{|l|r|r|r|r|r|r|}
\hline
\rowcolor[HTML]{DAE8FC} 
\textbf{} & \multicolumn{1}{l|}{\cellcolor[HTML]{DAE8FC}\textbf{$K$}} & \multicolumn{1}{l|}{\cellcolor[HTML]{DAE8FC}\textbf{$CRI$}} & \multicolumn{1}{l|}{\cellcolor[HTML]{DAE8FC}\textbf{$UGR$}} & \multicolumn{1}{l|}{\cellcolor[HTML]{DAE8FC}\textbf{$AVG$}} & \multicolumn{1}{l|}{\cellcolor[HTML]{DAE8FC}\textbf{$G1$}} & \multicolumn{1}{l|}{\cellcolor[HTML]{DAE8FC}\textbf{$G2$}} \\ \hline
add & 1 & 1 & 3 & 10 & 10 & 7 \\ \hline
remove & 1 & 1 & 10 & 1 & 5 & 5 \\ \hline
dim & 1 & 1 & 10 & 1 & 1 & 1 \\ \hline
height incr. & 1 & 1 & 10 & 4 & 6 & 10 \\ \hline
height decr. & 1 & 1 & 10 & 6 & 10 & 4 \\ \hline
change collection & 10 & 10 & 6 & 4 & 4 & 6 \\ \hline
change version & 10 & 10 & 1 & 1 & 6 & 4 \\ \hline
\end{tabular}
\end{table}

The WSM-score for each action is then given by
\begin{equation}
A^{\textup{WSM-Score}}_{i} = \sum_{j=1}^{n}w_{C_j} a_{ij}, \textup{for}~i = 1,2,3,...,m, 
\end{equation}
where $n$ is the number of modeling actions, $m$ is the number of criteria and $w_{C_j}$ is the importance of the criterion $C_j$ assigned using the weight sliders in the focus setting view (see \autoref{fig:weights}).
We select the two highest ranked modeling actions to generate potential solutions. For each action, we simulate between three to five randomized parameterizations, resulting in up to 10 simulated lighting designs.

To only present the three highest-scoring solutions to the lighting designer, we compute a progress score $s$ for each simulated scenario, and pick the top three options to be displayed as suggestions. The remaining solutions are filtered out and are not used. 
$s$ is a hierarchically calculated mean over fulfillment levels of the illumination constraints $C$ over all measurement objects $M$. 
First, the mean fulfillment level of a measurement object $\bar{f_M}$ per criteria is formed by:
\begin{equation}
    \bar{f_M} = \frac{\sum_{i=1}^{p} f_{M_i}}{p},
    \label{eq:progress}
\end{equation}
where $p$ is the number of measurement objects. 
Second, using the lighting designers assigned group weights $w_{G}$ (see \autoref{fig:weights}), the weighted average fulfillment level of a criteria over all groups $\bar{f_G}$ is defined by:
\begin{equation}
    \bar{f_G} = \frac{\sum_{i=1}^{o}(w_{G_i} * \bar{f}_{M_i})}{\sum_{i=1}^{o}w_{G_i}},
    \label{eq:constraint}
\end{equation}
where $o$ is the number of groups.
Finally, the score $s$ is composed by the weighted means $\bar{f_G}$:
\begin{equation}
   s = \frac{\sum_{i=1}^{m}(w_{C_i} * \bar{f_{G_i}})}{\sum_{i=1}^{m}w_{C_i}},
   \label{eq:score}
\end{equation}
where $m$ number of illumination constraints and $w_{C}$ is the weight of the illumination constraint.
For informative reasons, $s$ is presented to the designer in the top right corner of the quality view (see \autoref{fig:teaser} (d)). A value of 1.0 indicates a complete fulfillment of the objectives for the current weighting.
We observered that the randomized parameterization of the modeling tasks can lead to erroneous solutions (e.g., when a light moves through the ceiling during a height change), 
but they are filtered out automatically due to their low score. 


Our guidance tool targets the simplification of the tasks described in \autoref{sec:tasks}, and helps to evaluate a broader spectrum of possible modeling steps automatically. This decreases the time to converge to the fulfillment of the illumination constraints, allows for intuitive exploration of alternative solutions, and still supports manual intervention and artistic freedom. The guidance tool is designed to support the modeling process but is not intended to be a complete replacement for manual modeling.

\section{Implementation}

We implemented our proposed interactive visualization and guidance method in an existing lighting design software prototype with industry-standard modeling tools, a database with measured light sources, and an optimized, state-of-the-art GPU-based simulation kernel~\cite{PB-VRVis-2013-002}. The software was extended by integrating web-based visualizations (primarily created with \emph{D3}~\cite{Bostock2011}) using the Chromium browser libraries~\cite{chromium}, and were linked to the 3D view via websockets.
In our test scenes, a typical simulation run takes about 1 to 5 seconds to complete on our desktop workstation with an Intel i5 7660K CPU, 32GB RAM and a NVIDIA GTX 1080 graphics card at 1920x1080 resolution. The texel size of the light map corresponds to a real-world area of 1x1cm, and we limited the number of light bounces to three. The user study (see \autoref{sec:evaluation}) was conducted using the same system and settings. 

\section{Evaluation}
\label{sec:evaluation}
We conducted a qualitative user study with three lighting designers.
\subsection{Study Design}

We recruited three lighting designers as study participants to test our prototype. None of them was involved in the design process nor did they see our prototype before. We opted for a qualitative evaluation, which is effective when working with domain experts to better understand how they interact with the prototype, to observe which insights the experts can derive from the visualizations, to collect detailed feedback of the usefulness of the prototype, and how it fits to their workflow~\cite{isenberg:tvcg:2013, kriglstein:eurorv3:2015}.
We conducted individual testing sessions with one developer and the respective study participant. 

The sessions started with a short interview to learn about the evaluators' experience with visualizations and what tools they are currently using. We introduced LightGuider, its functionality, and explained the lighting scenario used for this session within 15 minutes. We encouraged the study participant to think aloud while performing the tasks mentioned below. The participants finished the tasks in approximately 20 minutes, without any further training. We then conducted a semi-structured interview to learn about their impressions of the prototype.

\begin{itemize}
    \item \textbf{Task 1: Identify needs for improvement. }
    Manipulate the pre-placed lights manually and use the bullet charts and provenance tree to reason if the lighting design changed for the better or worse (i.e., which illumination constraints are satisfied and which aspects require further attention?).
    \item \textbf{Task 2: Understand the provided guidance.}
    Use the provided guidance to improve the current lighting design. Generate new solutions and reason which one to choose and why.
    \item \textbf{Task 3: Set the focus to important design aspects} 
    Manipulate the weights that reflect how important the different illumination constraints are for different measuring surfaces.
    \item \textbf{Task 4: Compare lighting designs} 
    Select one lighting design of interest and use the comparison view and reason which lighting designs are superior or inferior in which aspects.
\end{itemize}

Task 1 was designed to see if lighting designers can efficiently interpret the provenance tree and the visualization of illumination constraints. Task 2 was aimed at evaluating if the provided guidance is understood and useful to the lighting designers. In Task 3 the designers should learn how to fine-tune the provided guidance as well as the visualization of the quality of the lighting designs. Finally, with Task 4 we wanted to understand if the comparison view helps to derive further insights. We encouraged all study participants to freely explore the prototype while executing the tasks and articulate any thoughts.
\subsection{Results}
\label{sec:results}
In this section, we summarize our observations and comments three study participants (referred to as P1, P2, and P3) phrased while interacting with the prototype, as well as in the subsequent interviews.
Our study participants are familiar with \textbf{interactive visualizations} to analyze lighting designs, however, measured data is mostly presented in form of static numeric tables.
\begin{quote}
    ``Great overview! Only one glance is enough to see everything, especially due to the visual presentation of the data in contrast to numeric tables.'' (P3)
\end{quote}

While interacting with the prototype, we could observe a common strategy among study participants to use the latest provenance tree node, i.e., the integrated 
\textbf{treemap} as a starting point to gain a rough overview of the quality of the current light design.
The lighting designers appreciated the color saturation-based quality encoding.

\begin{quote}
   ``After realizing that whitish areas represent fully satisfied requirements, it is really easy to interpret the current state.'' (P2) 
\end{quote}
\begin{quote}
    ``This one is the worst solution---I can see this because it contains the darkest colors. This one, on the other hand, is the best solution because it is lighter than the others.'' (P1)
\end{quote}
They used the summary-view layout (worst-case summary) of the treemap to gain a fast overview, and they expanded the nodes to access the detail-view in case of dark-colored treemap nodes to better understand which measuring surface does not meet the illumination constraints.
\begin{quote}
    ``I think the treemap is intuitive and easy to understand.'' (P2)
\end{quote}

When trying to improve the lighting design, they solely focused on more saturated colors and consulted the linked \textbf{bullet charts} for details.
One of the designers summarized the interaction with the bullet charts as follows: 
\begin{quote}
    ``You only have to check for the indicators [note: the black line indicating the measured value for the respective illumination constraint] to be within the white area on the right side to know it is satisfied, that is consistent and allows to detect non-satisfied constraints efficiently.'' (P3)
\end{quote}
They also appreciated the \textbf{weighting of constraints} for enabling a fast change of focus and for allowing them to efficiently filter out constraints that are not important in the given task or scenario. Besides using the linking and brushing functionality, they also used the consistently-applied color scheme a lot to relate the constraint sliders, bullet charts, and treemap visualizations. 
Furthermore, the designers made use of the linked \textbf{3D view} to inspect the scene setup in more detail and to identify the location of non-satisfied illumination constraints.
The \textbf{provenance tree} motivated the designers to experiment with alternative lighting design strategies to improve the current design. 
\begin{quote}
    ``I like the idea to go back to any state and to start a new branch.'' (P2)
\end{quote}
They could easily find the best design solution even in complex provenance trees after a long design session, and they really appreciated the visual means for comparing different solutions and aspects, which common tools do not provide in such an elegant way.
\begin{quote}
    ``With this tool the comparability is really good.'' (P1)
\end{quote}

\paragraph{Suggested Improvements.}
The lighting designers had a few improvement ideas, mostly related to usability, such as adding the possibility to remove nodes from the provenance tree to clean-up failed attempts
, to have the screenshot-strip only on demand
, or to hide the corresponding bullet charts (not only the treemap space) when assigning zero-weights to particular illumination constraints. 
However, they formulated interesting ideas for improving the guidance mechanism.
The first version of our guidance system took the current fulfillment level of illumination constraints into account, i.e., when a constraint was already satisfied the generated suggestions would not aim at further improvement, even if this constraint had a high assigned weight. However, our first study participant clearly rejected this solution. Thus, we adapted the guidance generation for the other two study participants so they could receive suggestions for further improving exactly the illumination constraint they set the focus on, even if this constraint was already met.
Moreover, all participants expressed the wish to fine-tune the suggestion generation, or as P2 phrased it: ``...\textbf{means to guide the guidance}''. While we did not yet include such a functionality, integrating mechanisms for designers to provide guidance themselves, and thus, to reveal the whole potential of guidance as a mixed-initiative process (as discussed by Ceneda et al.~\cite{ceneda2018guidance}), in which the guidance mechanism dynamically adjusts the provided suggestions to the designer's actions and feedback is an interesting research challenge stated by Ceneda et al.~\cite{mistery_1}. In particular, our study participants wanted to be able to set restrictions to limit the parameter space for suggested modeling steps, and thus, to steer the suggestions into a desired direction. For instance, lighting designers might want to use a particular type of luminaires for a given project, or luminaires should only be positioned at pre-defined locations. Still, guidance would be needed to optimize the remaining modeling parameters of these setups. 
Second, designers suggested to make the guidance mechanism \textbf{transparent} to the designer, which is related to the current research challenge of \emph{explainable AI} (artificial intelligence) presented by Samek et al.~\cite{DBLP:journals/corr/abs-1708-08296}. While they were excited about the potential of receiving automatically generated suggestions for design improvement, they still wanted to be better informed how the system came up with these suggestions and why.

\section{Discussion and Future Work}
\label{sec:discussion}
In this section we summarize gained insights as well as future plans.

\paragraph{Lessons Learned.}
While our qualitative user study showed that LightGuider was very well received by lighting designers, we also gained some interesting insights, in particular with respect to the provided guidance. Besides the two aspects we discussed in the previous section~\ref{sec:results}, i.e., the aspects of \textbf{steerable guidance} and \textbf{transparent guidance}, another interesting study result should be mentioned.
Our first study participant pointed out that it would be beneficial to receive suggestions exactly for the illumination constraint he set the focus on, even if it had already reached the minimum requirements---sometimes just for checking out alternative solutions that might be aesthetically more pleasing. He elucidated that while removing constraints makes sense from a pure optimization-problem perspective, and would be the right approach in pure mathematical/technical scenarios that do not have any creative/aesthetic aspects to him, it does not meet the real needs of light designers. This underlines the creative aspect of this process. Again, this finding is related to \emph{steerable guidance}. However, it also stresses the importance to understand the targeted users. This is known for visualization design, but it is also true for designing guidance. In our case, \textbf{user-tailored guidance} does not only need to support the optimization problem at hand, but also the creative freedom that is necessary for lighting design.

\paragraph{Generalizability.} 
Our presented interactive visualizations would very well lend itself to other application fields. The provenance tree in combination with the bullet charts for setting the subjective ``importance'' of a topic in combination with the automatically generated suggestions is a concept that fits decision support systems, tasks, such as data cleansing, or defining data quality metrics, or generally use cases in which interactions with a large, multi-dimensional data space can be recorded and guided. Our selected set of constraints, in the case of lighting design the illumination constraints, can easily be exchanged by any other measurement constraints. 

\paragraph{Limitations.} 
A shortcoming of our visualization approach in its current form is scalability. 
The number of parallel and sequential nodes in the provenance tree could pose a problem in longer modeling sessions. 
This can be solved by integrating a strategy for merging and folding of nodes (as demonstrated by Stitz et al.~\cite{stitz:CGF:2016}). 
Also, an increase of the number of illumination constraints would pose a challenge, because the number of well-distinguishable colors for their encoding is limited. 
We suggest to counter this problem by grouping these constraints. 
For significantly larger scenes with multiple rooms and measurement objects, the treemap leafs might get too small to be perceivable, which could be tackled by introducing another level of (semantic) hierarchy, and providing a provenance tree for each room. 
In terms of guidance, our approach scales well to the provision of more suggestions (e.g., computed in the cloud), as presenting only the best solutions makes sure that the lighting designer is not overwhelmed.

\paragraph{Future Research.}
Our provenance tree shares similarities to visualizations known from version control systems and we want to investigate which known operations from these tools (e.g., merging, ``cherry picking'', removal of actions, etc.) are applicable.
This would not only enable a more sophisticated, non-sequential undo mechanism (as shown for vector graphics editing by Su et al.~\cite{sara:mit:2009}), but would also allow designers to focus on solving local design problems that are then easily combined into a global solution---ultimately, even in multi-user setups. 
Besides, we believe that our approach could integrate lighting design methodologies of related works, e.g., the hierarchy-based weighting of objectives as well as the 3D comparisons proposed by Sorger et al.~\cite{sorger-2015-litevis}. In particular, the combination of their ranking approach with LightGuider's evaluation of parallel solutions could lead to a powerful, holistic lighting design approach. Using procedural light-source placement as presented by Schwarz et al.~\cite{Schwarz2014} for the generation of more advanced suggestion seems interesting as well, especially for the alignment of light sources added to the scene (currently, lights are added in the proximity of neighbors with a random offset).  
Finally, our current suggestion generation strategy is a comparatively simple heuristic based on typical actions from lighting designers to demonstrate the capabilities of our visual guidance tool. One could couple the generation of suggestions with the objects currently visible in the 3D view frustum, or improve the suggestions based on the tracked behavior of the designers (e.g., acceptance or rejection of suggestions, or manual modeling) in a machine-learning approach. Besides improving guidance suggestions, it could be a possibility to ``learn aesthetics''. The capabilities of our system offer the possibility to collect the large amounts of data needed for such learning approaches---maybe even leading to \emph{style transfer} in lighting design. 

\section{Conclusion}

We presented LightGuider, a prototype combining VA, 3D modeling, and simulation for enhancing interactive lighting design. 
One of our novelties is the integration of provenance visualizations and quality information as well as guidance to improve lighting design, while leaving all artistic freedom to designers.
Instead of a trial-and-error-based workflow leading to a single solution, LightGuider provides means to evaluate and explore different designs in a structured, comprehensible way. 
The provision of automatically generated suggestions for further improvement is essential to facilitate the laborious task of optimizing a design to fulfill the defined illumination constraints, and helps to discover alternative solutions. The enhanced provenance and lighting design quality visualizations enable the comparison of different design approaches and allow for an informed reasoning about their suitability.
Our qualitative evaluation with three designers showed that the participants very well perceived these features. They especially praised the great overview and visual comparison of modeling states and designs, the interactive visual feedback, and its potential to considerably facilitate and speed up their task of lighting design. ``This is a very valuable approach to make lighting design optimization easy and fast'' (P3).

\acknowledgments{We wish to thank Miriah Meyer, Thomas H\"{o}llt,  and Thomas Ortner for their insightful and valuable comments on this work, and the designers from Zumtobel Lighting for sharing their expert knowledge.
VRVis is funded by BMVIT, BMDW, Styria, SFG and Vienna Business Agency in the scope of COMET - Competence Centers for Excellent Technologies (854174) which is managed by FFG. 
This work is partly supported by the VIDI NextView, funded by NWO Vernieuwingsimpuls.

}

\bibliographystyle{abbrv-doi}

\bibliography{bibliography}

\begin{thebibliography}{10}

\bibitem{bors:eurovis:2018}
C.~Bors, T.~Gschwandtner, and S.~Miksch.
\newblock Visually exploring data provenance and quality of open data.
\newblock In {\em Proc. {EuroVis}}, pp. 9--11. Eurographics Association, Brno,
  Czech Republic, 2018.

\bibitem{Bostock2011}
M.~Bostock, V.~Ogievetsky, and J.~Heer.
\newblock D$^3$ data-driven documents.
\newblock {\em IEEE Transactions on Visualization and Computer Graphics},
  17(12):2301--2309, Dec. 2011.

\bibitem{Bouali:vc:2016}
F.~Bouali, A.~Guettala, and G.~Venturini.
\newblock {VizAssist: An} interactive user assistant for visual data mining.
\newblock {\em The Visual Computer}, 32(11):1447--1463, Nov. 2016.

\bibitem{colorBrewer}
C.~Brewer and M.~Harrower.
\newblock Color{B}rewer2.
\newblock http://colorbrewer2.org/.
\newblock Accessed: 2019-06-17.

\bibitem{brucker-2010-RES}
S.~Bruckner and T.~M\"{o}ller.
\newblock Result-driven exploration of simulation parameter spaces for visual
  effects design.
\newblock {\em IEEE Transactions on Visualization and Computer Graphics},
  16(6):1467--1475, Oct. 2010.

\bibitem{ceneda:tvcg:2017}
D.~Ceneda, T.~Gschwandtner, T.~May, S.~Miksch, H.-J. Schulz, M.~Streit, and
  C.~Tominski.
\newblock Characterizing guidance in visual analytics.
\newblock {\em IEEE Transactions on Visualization and Computer Graphics},
  23(1):111--120, Jan. 2017.

\bibitem{ceneda2018guidance}
D.~Ceneda, T.~Gschwandtner, T.~May, S.~Miksch, M.~Streit, and C.~Tominski.
\newblock {Guidance or No Guidance? A Decision Tree Can Help}.
\newblock In {\em Proc. {EuroVA}}, pp. 19--23. Eurographics Association, Porto,
  Portugal, 2018.

\bibitem{mistery_1}
D.~Ceneda, T.~Gschwandtner, and S.~Miksch.
\newblock A review of guidance approaches in visual data analysis: A multifocal
  perspective.
\newblock {\em Eurographics/IEEE VGTC Conference on Visualization (STAR)},
  38(3):forthcoming, June 2019.

\bibitem{CIE_117-1995-1}
{Commission Internationale de L'Eclairage: Standard CIE 117-1995 - Discomfort
  Glare in Interior Lighting}, 1995.

\bibitem{coffey2013design}
D.~Coffey, C.-L. Lin, A.~G. Erdman, and D.~F. Keefe.
\newblock Design by dragging: An interface for creative forward and inverse
  design with simulation ensembles.
\newblock {\em IEEE Transactions on Visualization and Computer Graphics},
  19(12):2783--2791, Dec. 2013.

\bibitem{dialux}
{DIAL GmbH}.
\newblock {DIALux}.
\newblock https://www.dial.de.
\newblock Accessed: 2019-06-17.

\bibitem{DIN_EN_12464-1}
{DIN Standard EN 12464-1: Light and lighting - Lighting of work places - Part
  1: Indoor work places}, Aug. 2011.

\bibitem{Elm2008}
N.~Elmqvist and P.~Tsigas.
\newblock {A taxonomy of 3d occlusion management for visualization}.
\newblock {\em IEEE Transactions on Visualization and Computer Graphics},
  14(5):1095--1109, Sept. 2008.

\bibitem{10.2307/168461}
P.~C. Fishburn.
\newblock Additive utilities with incomplete product sets: Application to
  priorities and assignments.
\newblock {\em Operations Research}, 15(3):537--542, May 1967.

\bibitem{chromium}
{Google LLC}.
\newblock {The Chromium Project}.
\newblock https://www.chromium.org/.
\newblock Accessed: 2019-06-17.

\bibitem{isenberg:tvcg:2013}
T.~{Isenberg}, P.~{Isenberg}, J.~{Chen}, M.~{Sedlmair}, and T.~{M\"oller}.
\newblock A systematic review on the practice of evaluating visualization.
\newblock {\em IEEE Transactions on Visualization and Computer Graphics},
  19(12):2818--2827, Dec. 2013.

\bibitem{kriglstein:eurorv3:2015}
S.~Kriglstein and M.~Pohl.
\newblock {Choosing the Right Sample? Experiences of Selecting Participants for
  Visualization Evaluation}.
\newblock In W.~Aigner, P.~Rosenthal, and C.~Scheidegger, eds., {\em Proc.
  {EuroRV3}}, pp. 23--25. Eurographics Association, Cagliari, Sardinia, Italy,
  2015.

\bibitem{PB-VRVis-2017-018}
K.~Kr\"osl, C.~Luksch, M.~Schw\"arzler, and M.~Wimmer.
\newblock Lite{M}aker: Interactive luminaire development using progressive
  photon tracing and multi-resolution upsampling.
\newblock In {\em Proc. {VMV}}, pp. 1--8. Eurographics Association, Bonn,
  Germany, 2017.

\bibitem{agi32}
{Lighting Analysts, Inc.}
\newblock {AGi32}.
\newblock http://agi32.com.
\newblock Accessed: 2019-06-17.

\bibitem{Lin2013}
W.-C. Lin, T.-S. Huang, T.-C. Ho, Y.-T. Chen, and J.-H. Chuang.
\newblock Interactive lighting design with hierarchical light representation.
\newblock {\em Computer Graphics Forum}, 32(4):133--142, July 2013.

\bibitem{PB-VRVis-2013-002}
C.~Luksch, R.~F. Tobler, R.~Habel, M.~Schw\"{a}rzler, and M.~Wimmer.
\newblock Fast light-map computation with virtual polygon lights.
\newblock In {\em Proc. {I3D}}, pp. 87--94. ACM, New York, NY, USA, Mar. 2013.

\bibitem{Marks97}
J.~Marks, B.~Andalman, P.~A. Beardsley, W.~Freeman, S.~Gibson, J.~Hodgins,
  T.~Kang, B.~Mirtich, H.~Pfister, W.~Ruml, K.~Ryall, J.~Seims, and S.~Shieber.
\newblock {Design Galleries: A} general approach to setting parameters for
  computer graphics and animation.
\newblock In {\em Proc. {SIGGRAPH}}, pp. 389--400. ACM Press/Addison-Wesley
  Publishing Co., New York, NY, USA, 1997.

\bibitem{Miksch2014}
S.~Miksch and W.~Aigner.
\newblock A matter of time: Applying a data\textendash{}users\textendash{}tasks
  design triangle to visual analytics of time-oriented data.
\newblock {\em Computers \& Graphics}, 38:286--290, Feb. 2014.

\bibitem{O'Donovan:CHI:2015}
P.~O'Donovan, A.~Agarwala, and A.~Hertzmann.
\newblock Design{Scape}: {Design} with {Interactive} {Layout} {Suggestions}.
\newblock In {\em Proc. {CHI}}, pp. 1221--1224. ACM, New York, NY, USA, 2015.

\bibitem{Okabe2007}
M.~Okabe, Y.~Matsushita, L.~Shen, and T.~Igarashi.
\newblock Illumination brush: Interactive design of all-frequency lighting.
\newblock In {\em Proc. {PG}}, pp. 171--180. IEEE Computer Society, Washington,
  DC, USA, 2007.

\bibitem{Ortner2016}
T.~Ortner, J.~Sorger, H.~Piringer, G.~Hesina, and E.~Gr\"{o}ller.
\newblock Visual analytics and rendering for tunnel crack analysis.
\newblock {\em The Visual Computer}, 32(6-8):859--869, June 2016.

\bibitem{Pellacini2007}
F.~Pellacini, F.~Battaglia, R.~K. Morley, and A.~Finkelstein.
\newblock Lighting with paint.
\newblock {\em ACM Trans. Graph.}, 26(2):Article 9, June 2007.

\bibitem{Pfister:2001:TFB:616070.618820}
H.~Pfister, B.~Lorensen, C.~Bajaj, G.~Kindlmann, W.~Schroeder, L.~S. Avila,
  K.~Martin, R.~Machiraju, and J.~Lee.
\newblock The transfer function bake-off.
\newblock {\em IEEE Computer Computer Graphics and Applications}, 21(3):16--22,
  May 2001. doi: {{%
10\hspace{.1pt}\discretionary{.}{%
}{.}\hspace{.4pt}1109\discretionary{/}{%
}{/}38\hspace{.1pt}\discretionary{.}{%
}{.}\hspace{.4pt}920623}}


\bibitem{ragan:tvcg:2016}
E.~Ragan, E.~Alex, J.~Sanyal, and J.~Chen.
\newblock Characterizing provenance in visualization and data analysis: An
  organizational framework of provenance types and purposes.
\newblock {\em IEEE Transactions on Visualization and Computer Graphics},
  22(1):31--40, Jan. 2016.

\bibitem{relux}
{Relux Informatik}.
\newblock {ReluxSuite}.
\newblock http://relux.com.
\newblock Accessed: 2019-06-17.

\bibitem{6280550}
H.~{Ribi\v{c}i\'{c}}, J.~{Waser}, R.~{Fuchs}, G.~{Bl\"{o}schl}, and
  E.~{Gr\"{o}ller}.
\newblock Visual analysis and steering of flooding simulations.
\newblock {\em IEEE Transactions on Visualization and Computer Graphics},
  19(6):1062--1075, June 2013.

\bibitem{SLE17}
N.~Z. Salamon, M.~Lancelle, and E.~Eisemann.
\newblock Computational light painting using a virtual exposure.
\newblock {\em Computer Graphics Forum}, 36(2):1--8, May 2017. doi: {{%
10\hspace{.1pt}\discretionary{.}{%
}{.}\hspace{.4pt}1111\discretionary{/}{%
}{/}cgf\hspace{.1pt}\discretionary{.}{%
}{.}\hspace{.4pt}13101}}


\bibitem{DBLP:journals/corr/abs-1708-08296}
W.~Samek, T.~Wiegand, and K.~M{\"u}ller.
\newblock Explainable artificial intelligence: Understanding, visualizing and
  interpreting deep learning models.
\newblock {\em ITU Journal: ICT Discoveries - Special Issue 1 - The Impact of
  Artificial Intelligence (AI) on Communication Networks and Services},
  1:1--10, Oct. 2017.

\bibitem{Schoeneman1993}
C.~Schoeneman, J.~Dorsey, B.~Smits, J.~Arvo, and D.~Greenberg.
\newblock Painting with light.
\newblock In {\em Proc. {SIGGRAPH}}, pp. 143--146. ACM, NY, USA, 1993.

\bibitem{Schwarz2014}
M.~Schwarz and P.~Wonka.
\newblock Procedural design of exterior lighting for buildings with complex
  constraints.
\newblock {\em ACM Trans. Graph.}, 33(5):166:1--166:16, Sept. 2014.

\bibitem{6876043}
M.~{Sedlmair}, C.~{Heinzl}, S.~{Bruckner}, H.~{Piringer}, and T.~{M{\"o}ller}.
\newblock Visual parameter space analysis: A conceptual framework.
\newblock {\em IEEE Transactions on Visualization and Computer Graphics},
  20(12):2161--2170, Dec. 2014.

\bibitem{Shesh2007}
A.~Shesh and B.~Chen.
\newblock Crayon lighting: Sketch-guided illumination of models.
\newblock In {\em Proc. {GRAPHITE}}, pp. 95--102. ACM, NY, USA, 2007.

\bibitem{simmhan_survey_2005}
Y.~L. Simmhan, B.~Plale, and D.~Gannon.
\newblock A survey of data provenance techniques.
\newblock {\em Computer Science Department, Indiana University, Bloomington
  IN}, 34(3):31--36, Sept. 2005.

\bibitem{SHPRALDEE18}
G.~Simons, S.~Herholz, V.~Petitjean, T.~Rapp, M.~Ament, H.~Lensch,
  C.~Dachsbacher, M.~Eisemann, and E.~Eisemann.
\newblock Applying visual analytics to physically based rendering.
\newblock {\em Computer Graphics Forum}, 38(1):197--208, July 2018. doi: {{%
10\hspace{.1pt}\discretionary{.}{%
}{.}\hspace{.4pt}1111\discretionary{/}{%
}{/}cgf\hspace{.1pt}\discretionary{.}{%
}{.}\hspace{.4pt}13452}}


\bibitem{sorger-2015-litevis}
J.~Sorger, T.~Ortner, C.~Luksch, M.~Schw\"{a}rzler, M.~E. Gr\"{o}ller, and
  H.~Piringer.
\newblock Lite{V}is: Integrated visualization for simulation-based decision
  support in lighting design.
\newblock {\em IEEE Transactions on Visualization and Computer Graphics},
  22(1):290--299, Jan. 2016.

\bibitem{stasko:2000}
J.~Stasko, R.~Catrambone, M.~Gzdial, and K.~McDonald.
\newblock An evaluation of space-filling information visualizations for
  depicting hierarchical structures.
\newblock {\em Int. J. Human-Computer Studies}, 53(5):663--694, Nov. 2000.

\bibitem{2018_vast_knowledge-pearls}
H.~Stitz, S.~Gratzl, H.~Piringer, T.~Zichner, and M.~Streit.
\newblock Knowledgepearls: Provenance-based visualization retrieval.
\newblock {\em IEEE Transactions on Visualization and Computer Graphics},
  25(1):120--130, Jan. 2019.

\bibitem{stitz:CGF:2016}
H.~Stitz, S.~Luger, M.~Streit, and N.~Gehlenborg.
\newblock {AVOCADO: V}isualization of workflow-derived data provenance for
  reproducible biomedical research.
\newblock {\em Computer Graphics Forum}, 35(3):481--490, June 2016.

\bibitem{sara:mit:2009}
S.~L. Su, S.~Paris, F.~Aliaga, C.~Scull, S.~Johnson, and F.~Durand.
\newblock Interactive visual histories for vector graphics.
\newblock Technical Report MIT-CSAIL-TR-2009-031, Massachusetts Institute of
  Technology (MIT), Cambridge, MA, June 2009.

\bibitem{viegas:chi:2004}
F.~B. Vi{\'e}gas, M.~Wattenberg, and K.~Dave.
\newblock Studying cooperation and conflict between authors with history flow
  visualizations.
\newblock In {\em Proc. {CHI}}, pp. 575--582. ACM, New York, NY, USA, 2004.

\bibitem{Waser2010}
J.~Waser, R.~Fuchs, H.~Ribicic, B.~Schindler, G.~Bl\"{o}schl, and
  E.~Gr\"{o}ller.
\newblock {World Lines}.
\newblock {\em IEEE Transactions on Visualization and Computer Graphics},
  16(6):1458--1467, Nov. 2010.

\bibitem{Yang:exn:2007}
D.~Yang, Z.~Xie, E.~A. Rundensteiner, and M.~O. Ward.
\newblock Managing discoveries in the visual analytics process.
\newblock {\em SIGKDD Explorations Newsletter}, 9(2):22--29, Dec. 2007.

\bibitem{lightinghandbook}
{Zumtobel Lighting GmbH}.
\newblock {The Lighting Handbook}.
\newblock https://www.zum- tobel.com/PDB/teaser/EN/lichthandbuch.pdf.
\newblock Accessed: 2019-06-17.

\end{thebibliography}

\end{document}